\documentclass[aps,prl,twocolumn,groupedaddress]{revtex4-1}
\usepackage{graphicx}
\usepackage{color}

\begin{document}

\title{Difference between level statistics, ergodicity and localization transitions on the Bethe lattice}

\author{G. Biroli$^1$}
\author{A.~C. Ribeiro Teixeira$^2$}
\author{M. Tarzia$^3$}
\affiliation{$^1$Institut de Physique Th\'eorique, CEA/DSM/IPhT-CNRS/URA 2306 CEA-Saclay,
F-91191 Gif-sur-Yvette, France}
\affiliation{$^2$Instituto de Física, Universidade Federal do Rio Grande do Sul, Caixa Postal 15051, CEP 91501-970, Porto Alegre, RS, Brazil}
\affiliation{$^3$LPTMC, CNRS-UMR 7600, Universit\'e Pierre et Marie Curie, bo\^ite 121, 4 Pl. Jussieu, 75252 Paris c\'edex 05, France}

\date{\today}

\begin{abstract}
We show that non-interacting disordered electrons on a Bethe lattice display a new intermediate phase
which is delocalized but non-ergodic, {\it i.e.}~it is characterized by Poisson instead of GOE statistics. 
The physical signature of this phase is a very heterogenous 
transport that proceeds over a few disorder dependent paths only. We show that the transition to the 
usual ergodic delocalized phase, which takes place for a disorder strength smaller than the one leading to 
the localization transition, is related to the freezing-glass transition of directed polymers in random media. 
The numerical study of level and eigenstate statistics, 
and of the singular properties of the probability distribution of the local density of states all support the existence of this new intermediate phase.
Our results suggest that the localization transition may change nature in high dimensional systems.  
\end{abstract}

\pacs{}

\maketitle

After more than a half century of Anderson localization, the subject is still very much alive~\cite{fiftylocalization} as proved by the recent 
observations of Anderson localization of atomic gases in one dimension~\cite{aspect} and of classical sound elastic waves 
in three dimensions~\cite{localizationelastic}. Also on the theoretical side several questions remain open: Although there is by now a good 
understanding of the localization transition in low dimensional systems, culminating in a functional renormalization group analysis by
a $2+\epsilon$ expansion~\cite{ludwig}, the behavior in high dimensions, in particular the existence of an upper 
critical dimension and the relationship with Bethe lattice analysis~\cite{abou}, is still an issue. 
Recently, there has been a renewal of interest on this problem because of its relationship with Many-Body localization, 
a fascinating phenomenon that should take place for disordered isolated interacting quantum systems and lead to a new kind of phase transition 
between a low temperature non-ergodic phase---a purely quantum glass---and a high temperature ergodic phase (a metal in 
the case of disordered electrons)~\cite{BAA}. A paradigmatic representation of this transition~\cite{A97,BAA} is indeed Anderson localization 
on a very high dimensional lattice---the Fock space---which for spinless electrons consists in an N-dimensional hypercube 
($N\gg1$ is the number of sites of the lattice system). 
Localization had an impact on several fields, in particular Random Matrices and Quantum Chaos. As a matter of fact, in the delocalized phase the level statistics is described by random matrix theory and generally corresponds to the Gaussian Orthogonal Ensemble (GOE), whereas instead 
in the localized phase is determined by Poisson statistics because wave-functions close in energy are exponentially localized on very distant sites and hence do not overlap; thus, contrary to the GOE case, there is no level-repulsion and eigen-energies are distributed similarly to random points thrown on a line. The relationship with quantum chaos goes back to the Bohigas-Giannoni-Schmidt conjecture, which states that the level statistics of chaotic (or ergodic) systems is given by random matrix theory, whereas integrable systems instead are characterized by Poisson 
statistics~\cite{BGS}. This result can be fully worked out and understood in the semi-classical limit~\cite{berry,altshulerchaos}: for a quantum chaotic 
system, in the $\hbar \rightarrow 0$ limit, wave-functions at a given energy become uniformly spread over the micro-canonical hyper-surface of the configuration space. They are fully delocalized as expected for an ergodic classical system that covers regions with same energy uniformly. Instead, quantum non-ergodic models, such as integrable systems, 
are characterized by Poisson statistics and localized wave-functions. All those results support a general  
relationship between delocalization--GOE statistics--ergodicity (similarly between localization--Poisson statistics--lack of ergodicity).
In this work, following up a suggestion in~\cite{A97}, we show that surprisingly non-interacting disordered electrons on a Bethe lattice instead display a new intermediate phase, which is delocalized and yet still not ergodic (henceforth ``not ergodic'' will be a synonym of ``characterized
by Poisson statistics''). 
Since the Bethe lattice should be representative of 
high dimensional 
lattices, our results suggest that for large enough dimensions
the localization transition may change nature and be characterized by a new intermediate phase, which actually 
could also play an important role in Many-Body localization~\cite{scardicchioMB}.    \\
The model we focus on consists in non-interacting spinless electrons in a 
disordered potential: 
\begin{equation} \label{eq:H}
{\cal H} = - t \sum_{\langle i,j \rangle} \left( c_i^{\dagger} c_j
+ c_j^{\dagger} c_i \right ) + \sum_{i=1}^N \epsilon_i c_i^\dagger c_i
\end{equation}
where the first sum runs over all the nearest neighbors couples of the
Bethe lattice, the second sum runs over all $N$ sites;
$c_i^\dagger$, $c_i$ are fermionic creation and annihilation operators,
and $t$ is the hopping kinetic energy scale, which we take
equal to $1$. The on-site energies $\epsilon_i$ are i.i.d. random variable uniformly
distributed in the interval $[-W/2,W/2]$. 
The Bethe lattice is defined as a $k+1$-random-regular-graph, 
{\it i.e.}~a graph
chosen uniformly at random among all graphs of $N$ sites where each
of the sites has connectivity $k+1$~\cite{wormald}. This lattice is like a Cayley tree 
wrapped onto itself~\cite{footnotetree}. \\
Localization on the Bethe lattice was first studied by Abou-Chacra, Anderson
and Thouless~\cite{abou} and then later by many others, see~\cite{efetof,fyodorov,monthus-garel,berkovits,aizenmann,semerjian,ourselves}
and refs.~therein. 
Many similarities, but also few important differences, with the $3d$ behavior have been found. 
They mainly concern the critical properties, which display exponential instead of power
law singularities~\cite{efetof,fyodorov}, and the Inverse Participation Ratio, defined as $\textrm{IPR} = \sum_{i=1}^N
| \psi_i |^4$ ($\psi_i$ is the value of the wave-function on site $i$), which was conjectured to have 
a discontinuous jump at the 
transition from a $O(1)$ toward a $1/N$ scaling~\cite{fyodorov}.
The level statistics was conjectured to display a transition from GOE to
Poisson concomitant with the localization transition.  
However, numerical studies didn't fully support this claim~\cite{berkovits}.
Moreover, arguments put forward in~\cite{A97} indicated that the two transitions could actually not coincide.\\
In the following we indeed unveil that Anderson localization on the Bethe lattice is more subtle than originally thought and that
some of the previous conjectures have actually to be revised. 
We focus on the $k=2$ case (connectivity three) and on the middle of the spectrum, 
$E=0$.  Previous works determined by studying transmission properties and dissipation propagation 
that the localization transition takes place at $W_L \simeq 17.5$~\cite{abou,monthus-garel,ourselves}.
In order to analyze the level statistics and clarify its relationship with the localization transition, we have diagonalized the Hamiltonian for several
system sizes $N = 2^n$, with $n=9,10,11,12,13$.
For each $N$, we have averaged over both the on-site quenched disorder, 
and on random graph realizations ($2^{20-n}$ samples).
Since we are interested in $E=0$, we only focused on $1/8$ of the eigenstates centered around  
the middle of the band (taking $1/16$ or $1/32$ of the states do
not affect the results). 
\begin{figure}
\includegraphics[scale=0.305,angle=-90]{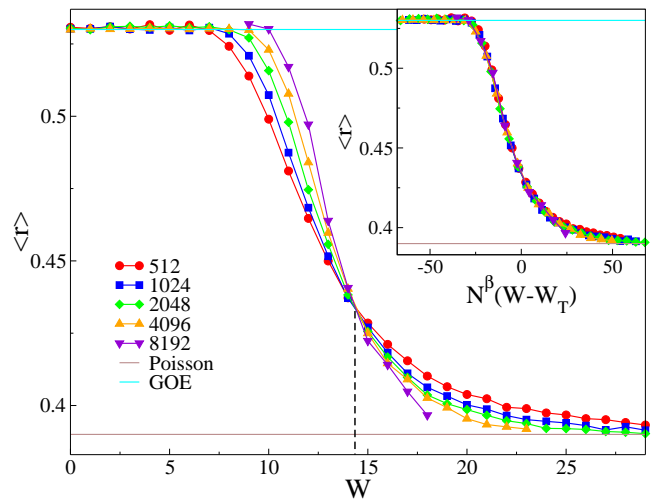}
\vspace{-0.2cm}
\caption{\label{fig:rav}
$\langle r \rangle$ as a function of $W$ for different system
sizes, showing that the transition from GOE to Poisson becomes sharper
as $N$ increases and all the curves cross at $W_T\simeq 14.5$. 
Inset: Finite size
scaling of the same curves showing data collapse ( 
$\beta \simeq 0.21$).}
\vspace{-0.4cm}
\end{figure}
We have studied the statistics of the level spacings of neighboring 
eigenvalues: $\delta_n^{(\alpha)} = E_{n+1}^{(\alpha)} - E_n^{(\alpha)} \ge 0$,
where $E_n^{(\alpha)}$ is the energy of the n-th eigenstate in the sample 
$\alpha$. In order to avoid problems related to the unfolding of the
spectrum, we measured the ratio of adjacent gaps~\cite{huse}:
$r_n^{(\alpha)} = 
\textrm{min} \{\delta_n^{(\alpha)},\delta_{n+1}^{(\alpha)}
\} / \textrm{max} \{\delta_n^{(\alpha)},\delta_{n+1}^{(\alpha)}
\}$, and obtained the (sample averaged) probability
distribution $P(r)$, which is expected to display a universal form 
depending on the level statistics~\cite{huse}. 
We did find that $P(r)$ converges to its GOE and Poisson counterparts 
for $W<W_T$ and $W>W_T$ respectively~\cite{footnoteWT} (the evolution with the system size of the entire distribution $P(r)$ is presented in the EPAPS). 
Surprisingly, however, $W_T$ turns out to be definitely smaller than $W_L$. 
In order to determine precisely $W_T$ we have focused on 
the finite size scaling behavior of $\langle r \rangle$.
As shown in Fig.~\ref{fig:rav}, the dependence of $\langle r \rangle$ on $W$ and $N$
can be described in terms of a scaling function of the variable $x=N^\beta (W-W_T)$, 
where $W_T\simeq 14.5$ and $\beta \simeq 0.21$. For $x\rightarrow -\infty$ and 
 $x\rightarrow +\infty$, we recover, as expected, the universal values $\langle r \rangle_{GOE} \simeq 0.53$ and 
$\langle r \rangle_P \simeq 0.39$ correspondingly to GOE and Poisson statistics. Fig.~\ref{fig:rav} clearly shows the existence of an intermediate phase for $W_T<W<W_L$. Independent evidence
can be gathered by studying the behavior of the average value of the IPR (averaged over samples and eigenstates close to $E=0$).  
Only for $W \lesssim 9$ we do find the standard scaling as $1/N$.
Between $9 \lesssim W \le W_T$, we find that $\langle \textrm{IPR} \rangle 
\sim 1/N^a$. The exponent $a \in (0,1]$ decreases
(almost linearly) from $1$ to $0$ and seems to vanish right at $W_T$.
In the new intermediate phase, $W_T \le W \le W_L$, the $\textrm{IPR}$
is better described by a logarithmic fit, $\langle \textrm{IPR} \rangle 
\sim 1/(\log N)^b$. The exponent $b$ decreases and apparently vanishes for 
$W \uparrow W_L$. 
Hence, wave-functions appear to be delocalized in the intermediate phase 
but on a number of sites so restricted that eigenstates are effectively 
independent, they do not repel each other and Poisson statistics holds. 
Complementary results, supporting our findings, have been obtained in~\cite{scardicchio}.\\
The previous analysis was performed for finite lattices. In order to overcome
this limitation and obtain a numerically exact solution 
directly for an infinite lattice, we follow Abou-Chacra {\it et al.}~\cite{abou} and analyze 
the recursive equations satisfied by the resolvent matrix $G(z) =  (H - z {\cal I}  )^{-1}$ on a Bethe lattice:
\begin{equation} \label{eq:recursion}
G_{i \to j}^{-1} (E + i \eta) = \epsilon_i - E - i \eta - \!\! 
\sum_{j^\prime \in
\partial i / j} \!\! G_{j^\prime \to i} (E + i \eta),
\end{equation}
where $\partial i/j$ denotes the set of neighbors of $i$ but $j$, and $G_{i \to j}^{-1}$
is the diagonal element of the resolvent matrix on site $i$ for a modified lattice 
where the edge between site $i$ and $j$ has been removed. 
A similar equation, where the sum on the LHS is over {\it all} neighbors of $i$ allows
one to get the local resolvent $G_{ii}$ and, from it, spectral properties such as
the global density of states, $\rho(E) = (1/N) \sum_{\alpha} \delta(E - E_{\alpha})=\lim_{\eta \to 0} (1/N \pi) \sum_i \textrm{Im} G_{ii}$,
and $\langle \textrm{IPR}\rangle= \lim_{\eta \to 0} \eta | G_{ii} (E + i \eta)|^2$.
Since the $G_{i \to j}$s are random variables, Eq.~(\ref{eq:recursion}) leads to a self-consistent 
equation on their probability distribution,  $P(G_{i \to j})$, which can be solved numerically 
using a population dynamics algorithm~\cite{abou,monthus-garel,ourselves}. 
In our analysis we used a population of ${\cal M} = 2^{26}$ elements.
Note that the value of ${\cal M}$ needed to correctly sample the tails of the distributions increases when
$\eta$ decreases, for ${\cal M} = 2^{26}$ we could only access  $\eta$ in the range
$10^{-2}-10^{-6}$.  \\
\begin{figure}
\includegraphics[scale=0.305,angle=-90]{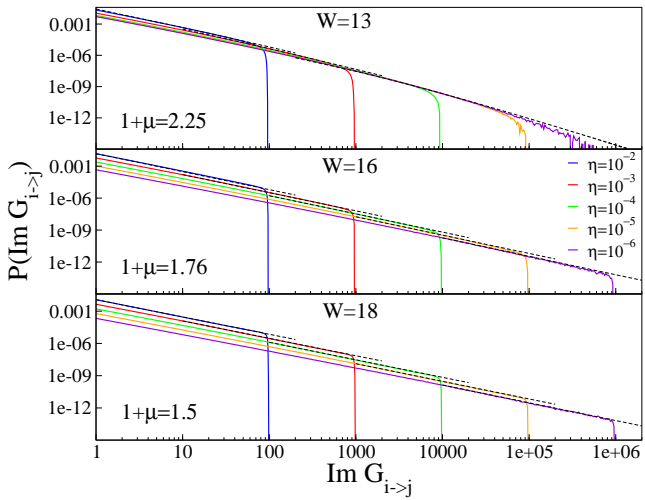}
\vspace{-0.2cm}
\caption{\label{fig:tails}
Power-law tails of the probability distributions of the imaginary part
of $G_{i \to j}$ in the delocalized ergodic phase ($W=13$, top panel), 
delocalized non-ergodic intermediate phase ($W=16$, middle panel), and
insulating localized phase ($W=18$, bottom panel), showing the value
of the exponent $\mu$. Note that only the curves in the top panel show 
a convergences to a $\eta$-independent value.}
\vspace{-0.4cm}
\end{figure}
We found that in the localized phase, $W>W_L$, 
the probability distribution of the imaginary part of $G_{i \to j}$ 
is singular in the limit $\eta \to 0$. Almost all values of $\textrm{Im} G_{i \to j}$
are of order $\eta$ except very rare ones.
More precisely, in the small $\eta$ limit, $P(\textrm{Im} G_{i \to j})$
has a scaling form $f(x/\eta)/\eta$ for $x\sim \eta$, where $\int dy f(y)=1$, 
and very fat tails: 
\begin{equation} \label{eq:tail}
P(\textrm{Im} G_{i \to j}) \sim \frac{c \, \eta^{1-\mu}}{(\textrm{Im} G_{i \to j})^{1+\mu}},
\end{equation}
with $c$ being a constant $O(1)$, and the exponent 
$\mu=1/2$ in the whole localized phase. The power law behavior is cut-off 
for $\textrm{Im} G_{i \to j}\sim 1/\eta$, see bottom panel of Fig.~\ref{fig:tails}. 
Note that since $\mu<1$ the tails give the leading contribution to the DOS, 
whereas the bulk part only a vanishing one. 
The probability distribution of the real part of $G_{i \to
j}$ instead converges for $\eta \to 0$ to a stationary 
distribution with power law tails with exponent $\mu = 1$. 
For $W<W_T$ we find that as $\eta \to 0$ the whole $P(G_{i \to j})$
has a non-singular $\eta$-independent limit characterized by power law tails (with logarithmic corrections) with an exponent $\mu>1$, see the top panel of Fig.~\ref{fig:tails} (and the EPAPS for more details). 
All these findings are in agreement with previous results. 
What is surprising compared to previous expectations is that there is not a direct transition between
these two regimes but instead, for $W_T<W<W_L$, we find an intermediate phase 
characterized by a mixed behavior of $P(\textrm{Im} G_{i \to j})$ and $P(\textrm{Re} G_{i \to j})$, 
that have a non-singular $\eta$-independent limit~\cite{aizenmann,fyodorov}
but, at the same time, display singular power law tails such as~(\ref{eq:tail})
(likely with additional logarithmic corrections)
 with an exponent stuck to $1$ for the real part 
and changing from $1/2$ to $1$ for the imaginary part, see the middle panel of Fig.~\ref{fig:tails}~\cite{footnote1}. 
Note that all the results we presented for $G_{i \to j}$
also hold for $G_{ii}$.
\begin{figure}
\includegraphics[scale=0.295,angle=-90]{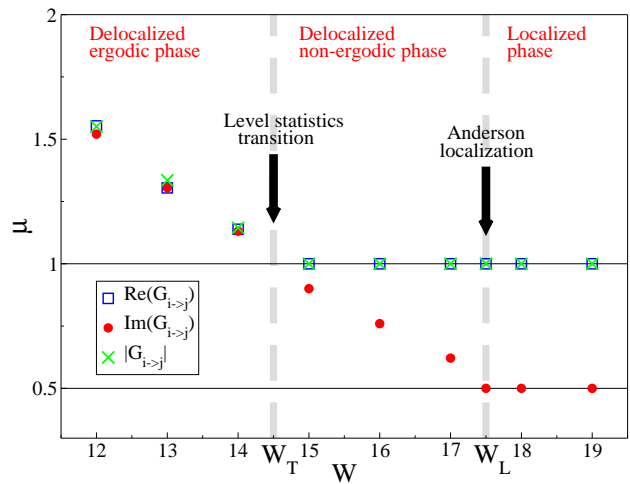}
\vspace{-0.2cm}
\caption{\label{fig:exp}
Exponent $\mu$ of the power law tails of the probability distribution
of the imaginary part, of the real part, and of the modulus of
$G_{i \to \j}$ as a function of $W$, showing the two transitions
at $W_T$ and $W_L$. }
\vspace{-0.4cm}
\end{figure}
The evolution of the exponent $\mu$, which is reported in Fig.~\ref{fig:exp}, clearly shows  
that a transition, different from Anderson localization, takes place at $W_T$.
Since $\lim_{\eta \rightarrow 0}\textrm{Im} G_{ii}= \pi \sum_{n} \delta(E - E_n^{(\alpha)} )|\psi_{n,i}^{(\alpha)}|^2$ 
one expects that the rare and very large values of $\textrm{Im} G_{ii}$ are due to similar rare events in 
$|\psi_{n,i}^{(\alpha)}|^2$. This is confirmed by exact diagonalization of finite systems (see EPAPS): 
the distribution of $|\psi_{n,i}^{(\alpha)}|^2$ displays power law tails with the {\it same} exponent $\mu$ appearing in~(\ref{eq:tail}). 
More results, including multi-fractal behavior, will be presented elsewhere~\cite{futurepaper}.\\
The nature of the transition taking place at $W_T$ can be understood by a mapping to directed polymers in random media (DPRM)~\cite{derrida}.
In order to do this, we telescope the recursive equation on $\textrm{Im} G_{i \to j}$ and end up with the expression~\cite{aizenmann}:
\begin{equation} \label{eq:DP}
\textrm{Im} G_{i \to j} = \sum_{{\cal P}}
\prod_{(i^\prime,i^{\prime \prime})\in {\cal P}} \! |G_{i^\prime \to 
i^{\prime \prime}}|^2 \, \textrm{Im}  G_{i_{R} \to i_{R-1}},
\end{equation}
where ${\cal P}$ are all the directed paths of length $R$ of the (rooted) 
Bethe lattice originating from site $i$, and $(i^\prime,i^{\prime \prime})$ 
are all the edges, including $i \to j$, belonging to the path ${\cal P}$. (We neglected the $\eta$-contribution 
since for $W_T<W<W_L$ the typical values of $\textrm{Im}  G_{i \to j}$ are of the order of one
whereas $\eta \rightarrow 0$.) 
The sum in~(\ref{eq:DP}) is over an exponential number of paths, $k^R$; in the large $R$ limit,
two cases are possible: the sum is dominated either by few paths that give a $O(1)$ contribution
or by an exponential number of paths, $k'^R$, each of them giving a very small contribution but such that their sum is $O(1)$
($k'$ is less than $k$ and $W$-dependent). We found that the transition between these two regimes happens precisely at $W_T$ and it is related to the glass transition of DPRM. 
Indeed, by introducing the edge-energy, $\omega_{i^{\prime} \to i^{\prime \prime}}$, by the equation $e^{-  
\omega_{i^{\prime} \to i^{\prime \prime}}}
= |G_{i^{\prime} \to i^{\prime \prime}}|^2$, one can re-interpret  
$\sum_{{\cal P}}
\prod_{(i^\prime,i^{\prime \prime})\in {\cal P}} |G_{i^\prime \to 
i^{\prime \prime}}|^2$ as the partition function for a DP on the Bethe lattice in presence of quenched bond disorder.
This problem can be solved, even in the case of disorder correlated as the $|G_{i^{\prime} \to i^{\prime \prime}}|^2$s~\cite{derrida,monthus-garel,futurepaper}, by computing the generalized ``free-energy'' (also introduced in~\cite{aizenmann}):
\begin{equation} \label{eq:phi}
\phi (s) = \lim_{\eta \to 0} \lim_{R \to \infty} \frac{1}{Rs}
 \overline{\log \left(\sum_{{\cal P}} \!
\prod_{(i^\prime,i^{\prime \prime})\in {\cal P}} |G_{i^\prime \to
i^{\prime \prime}}|^s \right)},
\end{equation}
It is easy to check that $\phi(s)$ is a convex function and that 
$\phi(s=2)=0$ for $W<W_L$ since typical values of $\textrm{Im} 
G_{i \to j}$ are of the order of one.  The average free energy of the directed polymer
has a freezing-glass transition, called one-step Replica Symmetry Breaking, 
akin to the one of the Random Energy Model~\cite{derrida,rem}: 
by decreasing the ``temperature'' $1/s$ the generalized free energy decreases until the critical point $s^\star$, 
defined by $\partial \phi(s) / \partial s |_{s = s^{\star}} = 0$, is reached;
for $s>s^*$ the DP freezes
 and its free energy remains constant: $\phi(s)=\phi(s^\star)$. 
In this glass phase the number of paths contributing to~(\ref{eq:phi}) is 
not exponential in $R$, as it happens for $s<s^\star$, but instead $O(1)$.
It has been rigorously proved in~\cite{aizenmann} (and indirectly found in~\cite{abou}, see~\cite{monthus-garel,futurepaper} for a detailed explanation) that the localization transition 
takes place for $s^\star=1, \phi(s^{\star})=0$~\cite{footnoteAi}. By 
computing $\phi(s)$ using recursive equations we have indeed found that
$s^\star=1$ at $W=W_L$; when diminishing $W$ below $W_L$ 
the value of $s^*$ increases and eventually reaches $2$ for $W=W_T$.
(see the EPAPS for details and plots of $\phi(s)$). 
In consequence, the mapping on DPRM unveils that the intermediate phase corresponds to a regime in which 
there is delocalization ($s^\star>1$), but the sum in~(\ref{eq:DP}) is dominated by few paths only ($s^\star<2$).
The statistical behavior of $\textrm{Im} G_{i \to j}$ can also be understood in terms of DPRM: it was shown in~\cite{derrida} that in the freezing-glass phase the DP partition function, and hence 
$\textrm{Im} G_{i \to j}$, display power law tails with an exponent $s^\star/2$ and logarithmic corrections;
this explains the emergence of the exponent $\mu$ that we indeed find equal to $s^\star/2$. Note, in particular, that
$s^\star=1$ and $s^\star=2$ correctly match the values $\mu=1/2$ and $\mu=1$ at $W=W_L$ and $W=W_T$.\\ 
In conclusion we have shown that disordered non-interacting electrons on a Bethe lattice 
display a new phase that is delocalized and non-ergodic. This comes to a large extent as
a surprise for a model that was considered to be exactly solved~\cite{abou,efetof,fyodorov}.
Our results are not actually in contradiction with previous ones since those mostly focused on 
the localization transition at $W_L$. The only disagreement regards the behavior of the 
$\langle \textrm{IPR}\rangle$ that was supposed to behave as $1/N$ in the whole delocalized phase~\cite{fyodorovsparse}. 
This is likely due to a hypothesis of regularity of the order parameter function intervening 
in the solution~\cite{fyodorov} that does not hold~\cite{footnote2}.
The mapping to DPRM unveils the peculiar nature of the delocalized non-ergodic phase: 
the level broadening for a given site $i$, $\textrm{Im} G_{ii}$, is finite for 
$\eta \to 0$, but cutting an arbitrarly far away bond chosen on one of the few paths contributing to~(\ref{eq:DP})  
substantially changes $\textrm{Im} G_{ii}$. Physically, this means that dissipation and transport
are very heterogenous: particles can travel far away but on a few and very specific, disordered dependent, paths. 
 Bethe lattice results are often considered representative of the infinite
dimensional limit. In consequence our results suggest that for large enough dimensions
the localization transition may change nature. This possibility, the implications for Many Body localization, 
the relationship with similar intermediate phases~\cite{chamon,pierre,levy} and the connection with DPRM~\cite{somoza,ioffe,markus,castellani} certainly are worth further investigations.   

\begin{acknowledgments}
We thank I. Aleiner, B. Altshuler, J.-P. Bouchaud, C. Castellani, Y. Fyodorov, T. Garel, P. Le Doussal, C. Monthus, M. Muller, V. Oganesyan, A. Scardicchio, G. Semerjian, S. Warzel for useful discussion and (GB)
ANR FAMOUS for support. We thank J.-P. Bouchaud, T. Garel, G. Semerjian and F. Zamponi 
for a careful reading of the manuscript.
\end{acknowledgments}

\appendix
\section{EPAPS}
In this supplementary material we provide more details and results related to several points discussed in the main text. 
\begin{figure}
\includegraphics[scale=0.305,angle=-90]{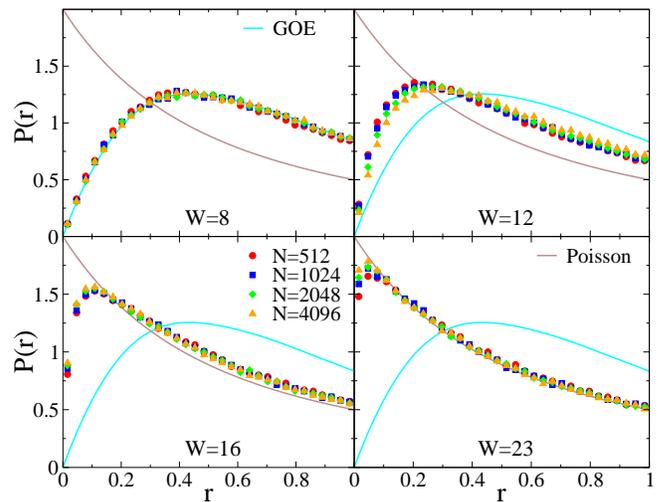}
\vspace{-0.2cm}
\caption{\label{fig:Pr}
Probability distribution of the average gap ratio for different system
sizes (from $N=512$ to $N=4096$) 
and for four different values of the disorder. The Poisson and GOE counterparts of $P(r)$ are also shown.
Top-left panel: Weak disorder, $W=8$.
The entire probability distribution $P(r)$ is described by the GOE
ensemble. Top-right panel: Moderately weak disorder, $W=12$. $P(r)$ is approaching the
GOE distribution as the system size is increased. Bottom-left panel: Moderately strong
disorder, $W=16$. $P(r)$ is converging towards the Poisson distribution
as $N$ is increased. Bottom-right panel: Strong disorder, $W=23$. The entire probability
distribution is described by the Poisson one, except at very low $r$,
where convergence is exponentially slow due to finite size effects.}
\vspace{-0.4cm}
\end{figure}

\subsection{Level statistics: The probability distribution of the gap ratio.}
In the case of Poisson statistics the probability distribution of the
gap ratio
$r$ is given by $P (r) = 2/(1 + r)^2$ (brown curve of Fig.~\ref{fig:Pr}), and its mean value is $\langle r \rangle_P
= 2 \log 2 - 1 \simeq 0.386$. For very strong
disorder, in the localized regime,
the numerical results show
that $P(r)$ has indeed this form (except for very small values of $r$,
for which the data approach the Poisson distribution exponentially slow
due to finite size effects), as shown in the bottom-right panel of Fig.~\ref{fig:Pr} 
for $W=23$.
For intermediate strong disorder, 
full convergence is not reached for our biggest system size. Nevertheless,
it is clear that $P(r)$ is converging toward
the Poisson distribution as $N$ is increased, as shown in the bottom-left 
of Fig.~\ref{fig:Pr} for $W=16$.

No analytical expression for the probability distribution of the gap ratio is known
in the case of GOE statistics but it 
can be easily determined numerically from exact diagonalization of
large GOE random matrices (cyan curve of Fig.~\ref{fig:Pr}). The mean value
$\langle r \rangle_{GOE}$ is equal to $0.5295 \pm 0.0006$. Level
repulsion in the GOE spectra manifests itself in the vanishing of the
probability distribution at $r=0$.
As expected, for weak disorder, the entire probability distribution
$P(r)$ is described by GOE, as shown in the top-left panel of Fig.~\ref{fig:Pr} for $W=8$.
For intermediate low disorder, although full convergence is not reached for the biggest sistem size,
$P(r)$ is converging toward the GOE distribution as $N$ is increased, as shown in the top-right panel of
Fig.~\ref{fig:Pr} for $W=12$.
Right at $W_T$ $P(r)$ is described by a stationary $N$-independent 
distribution (not shown), at least for the range of system sizes that we considered,
which does not correspond neither to GOE nor to Poisson statistics 
and should be universal.

These findings provide numerical evidence for the transition in the level statistics
from the GOE to the Poisson universality class. Such transition does
not take place at the Anderson localization transition, $W_L$,
as argued in~\cite{berkovits}, but
at a smaller value of the disorder, $W_T$.
Consider, for instance,
what happens for $W=16<W_L$: although the system
is in the delocalized phase, $P(r)$ is converging towards the Poisson
distribution and $\langle r \rangle$ is approaching $\langle r \rangle_P$
as $N$ is increased (as we show by finite size scaling in the main text).

\begin{figure}
\includegraphics[scale=0.29,angle=-90]{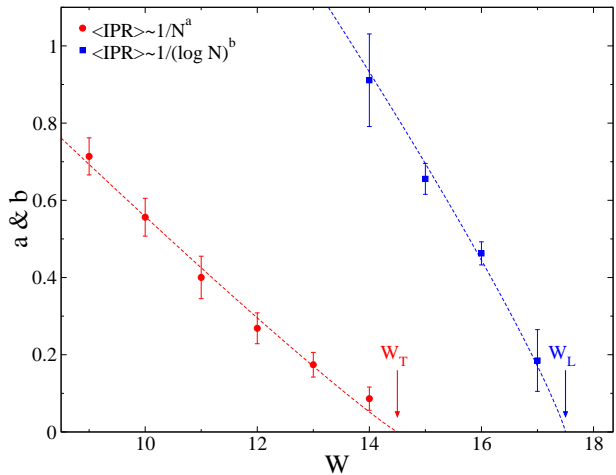}
\vspace{-0.2cm}
\caption{\label{fig:iprexp} $W$ dependence of the exponents $a$ and $b$ describing the behavior of $\langle \textrm{IPR}
\rangle$ as a function of the system size.}
\vspace{-0.1cm}
\end{figure}

\begin{figure}
\includegraphics[scale=0.31,angle=-90]{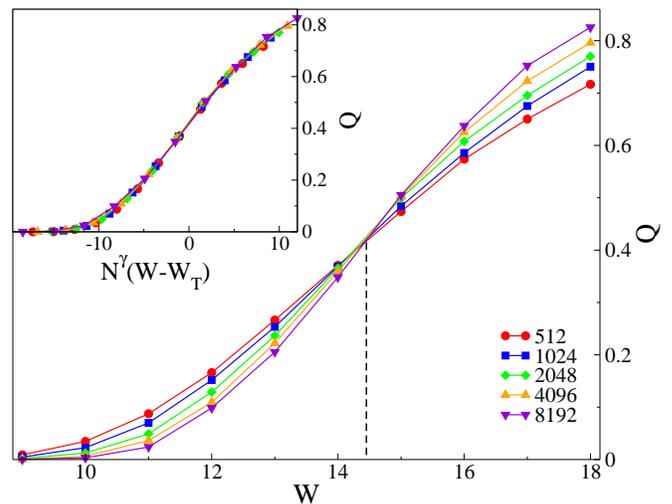}
\vspace{-0.2cm}
\caption{\label{fig:ipr} 
Main panel: $Q = \textrm{prob} ( \textrm{IPR}_n^{(\alpha)} > c
/ \log N) $
as a function of the disorder $W$ for different system sizes, and for $c=1.5$.
$Q$ grows from $0$ to $1$ as the disorder is increased. The transition becomes
sharper for bigger sizes.
The curves obtained for different system sizes all cross around $W \simeq W_T$.
Inset: Finite size scaling plot of $Q$ as a function of the variable
$N^\gamma (W -W_T )$ for different system sizes showing data collapse 
for $W_T = 14.5$ and $\gamma = 0.14$ 
(the results should be independent of $c$ for $N$ large enough, 
however for the limited range of $N$ at our disposal $c\simeq 1.5$ provides the best crossing).
}
\vspace{-0.4cm}
\end{figure}

\subsection{Inverse Participation Ratio.}
The $\textrm{IPR}$ of the $n$-th eigen-function of sample $\alpha$ is
defined as $\textrm{IPR}_n^{(\alpha)} = \sum_{i=1}^N |\psi_{n,i}^{(\alpha)}
|^4$, where $\psi_{n,i}^{(\alpha)}$ is the value of eigen-state $n$ of
sample $\alpha$ on site $i$.
In Fig.~\ref{fig:iprexp} we plot the values of the exponents $a$
(red circles)
and $b$ (blue squares) describing the system size
dependence of $\langle \textrm{IPR}
\rangle$ (averaged over eigen-functions around $E=0$ and over all
samples) as $\langle \textrm{IPR} \rangle \sim 1/N^a$ and 
$\langle \textrm{IPR} \rangle \sim 1/(\log N)^b$ respectively, as
a function of the disorder.
In the delocalized ergodic phase, $W<W_T$, $\langle \textrm{IPR} \rangle
\sim c/N^a$, with the exponent $a$ decreasing from $1$ at
$W\simeq 9$ to $0$ at $W=W_T$ as the disorder is increased.
In the intermediate delocalized non-ergodic phase, $W_T < W < W_L$,
the average $\langle \textrm{IPR} \rangle$ is better described by a logarithmic behavior,
$\langle \textrm{IPR} \rangle
\sim c/ (\log N)^b$. The exponent $b$ decreases as the disorder is increased
and vanishes at $W_L$. Finally, for $W>W_L$, $\langle \textrm{IPR} \rangle$
goes to a constant of $O(1)$ as $N$ is increased.
 
In order to rationalize the unexpected behavior of the average $\textrm{IPR}$
in the intermediate delocalized non-ergodic phase, we have analyzed the quantity
$Q = \textrm{prob} (\textrm{IPR}_n^{(\alpha)} > c/\log N)$, defined as
the probability that, for a given eigen-function $n$ of a given sample $\alpha$,
$\textrm{IPR}_n^{(\alpha)}$ is bigger than $c/\log N$ ($c$ being a constant of order $1$)
averaged over eigen-states close to $E=0$ and over all samples.
If the average $\textrm{IPR}$ behaves as a power law as a function of the system size, 
$\langle \textrm{IPR} \rangle \sim 1/N^a$ (as in the fully delocalized
phase) then $Q$ must vanish
in the thermodynamic limit. On the contrary, if either $\langle \textrm{IPR} \rangle$ goes
to a constant of $O(1)$ for large sizes (as in the fully localized phase), or
$\langle \textrm{IPR} \rangle \sim 1/(\log N)^b$ with $b<1 $(as in the intermediate phase),
then $Q$ must approach $1$ as $N \to \infty$. 
We indeed observe that $Q$ grows from $0$ to $1$ as the disorder is increased 
and that this transition becomes sharper as the system size grows.
For a suitable choice of the constant $c=1.5$, 
the curves obtained for different system sizes all cross at
$W \simeq W_T$ and
can be described in terms of a scaling function of  
the variable $N^\gamma (W - W_T)$, where $W_T \simeq 14.5$ and $\gamma = 0.14$, 
see Fig.~\ref{fig:ipr}.
The value of $\gamma$ is a bit different from the exponent used to collapse the $\langle r \rangle$
data. However, because of the rather limited range of system sizes and the small value of the exponents 
we do not consider this difference meaningful. Much larger system sizes would be necessary to 
get a precise estimate of the exponents.   

\begin{figure}
\includegraphics[scale=0.305,angle=-90]{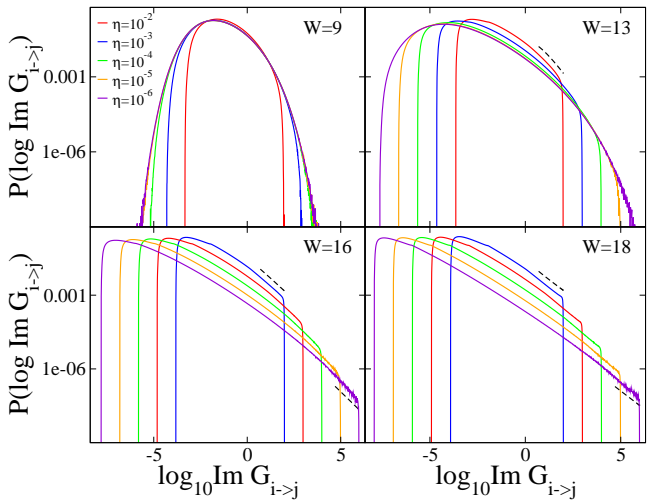}
\vspace{-0.2cm}
\caption{\label{fig:limits} 
Full probability distributions of $\log \textrm{Im} G_{i \to j}$ for
different values of the imaginary cut-off $\eta$ from $10^{-2}$ to $10^{-6}$
for four different values of the disorder. Top-left panel: $W=9$. 
$P (\log \textrm{Im} G_{i \to j})$ converges to a stationary 
$\eta$-independent distribution. Top-right panel: $W=13$. 
$P (\log \textrm{Im} G_{i \to j})$ is approaching a stationary
distributions although we are not able to observe full convergence for
the smallest allowed values of $\eta$. The tails of the distributions are described
by a power law (with logarithmic corrections) with an exponent $\mu = 1.3$
(black dashed line).
Bottom-left panel: $W=16$. In the intermediate delocalized non-ergodic phase
$P (\log \textrm{Im} G_{i \to j})$ is bound to have a non-singular
behavior for $\eta \to 0$, as proven in~\cite{aizenmann}. However, we are
apparently still very far from convergence. The tails of the distributions are
described by a singular power law (with logarithmic corrections) 
with an exponent $\mu = 0.76$ (black dashed lines), whose coefficient
behaves as $\eta^{1-\mu}$. 
Bottom-right panel: $W=18$. In the localized
phase the limit $\eta \to 0$ is singular. Almost all $\textrm{Im}  
G_{i \to j}$ are of order $\eta$, except very rare ones described 
by a power law (with logarithmic corrections) with an exponent $\mu = 1/2$
(black dashed lines),
whose coefficient vanishes as $\sqrt{\eta}$.} 
\vspace{-0.4cm}
\end{figure}

\subsection{Singular behavior of the local density of states.}
In Fig.~\ref{fig:limits} we show the full probability distributions of $\log(\textrm{Im}  G_{i \to j})$
for several values of the cut-off $\eta$.
In the fully localized phase, $W>W_L$, the probability distribution of the imaginary part of $G_{i \to j}$
has a singular behavior for $\eta \to 0$, as shown in the bottom-right panel of Fig.~\ref{fig:limits} for $W=18$. 
Almost all the values of $\textrm{Im}  G_{i \to j}$ are 
of order $\eta$, except for very rare events (whose fraction vanishes as $\eta$), 
described by very fat power law tails with an exponent $1 + \mu = 1.5$ 
(see Eq.~(\ref{eq:tail}) and Fig.~\ref{fig:tails}). 
Such tails give a contribution
of $O(1)$ to the local density of states, whereas
the bulk part only yields a vanishing contribution.
The probability distribution of the real part of $G_{i \to j}$ (not shown)
converge to a stationary distribution with power law tails with
a $W$-independent exponent $1+\mu = 2$.
 
For $W<W_L$, since the system is delocalized and the spectrum is absolutely continuous, $P (\textrm{Im}  G_{i \to j})$
must have a non-singular limit as $\eta \to 0$, as shown rigorously in~\cite{aizenmann}.
Indeed, for low enough disorder the whole $P(\log(\textrm{Im} G_{i \to j}))$ converges for $\eta \to 0$ to
a stationary non-singular $\eta$-independent distribution, as shown in the top-left panel of Fig.~\ref{fig:limits} for $W=9$.
This is also the case for $W=13$ (top-right panel of Fig.~\ref{fig:limits}), 
although we are not able to observe full convergence
for the smallest value of $\eta$ we are allowed to consider ($\eta\gg 1/\mathcal{M}$).
The tails of $P (\textrm{Im} G_{i \to j})$ are described by a power law (with logarithmic corrections) with an exponent
$1 + \mu$ with $\mu > 1$.
The probability distribution of the real part (not shown) also converges
to a stationary distribution described by power law tails with the
same exponent.

Finally, in the bottom-left panel of Fig.~\ref{fig:limits} we show the behavior of $P(\log(\textrm{Im}  G_{i \to j}))$ 
in the intermediate delocalized non-ergodic phase. The rigorous result of~\cite{aizenmann} guarantees us that   
the limit $\eta \to 0$ is not-singular and that a stationary distribution exists. However, 
for the smallest value of $\eta$ that we are allowed to consider we do not observe any convergence yet. 
The most likely reason is that since typical values of $\textrm{Im} G_{i \to j}$ are finite (for $\eta \to 0$) but extremely small close to $W_L$, 
(they vanishes as $\exp(c/\sqrt{W_L-W})$~\cite{fyodorov}), values of $\eta$ much smaller than the ones we can use
are needed to observe convergence. We are instead able to observe in full detail that the distribution 
of $\textrm{Im}  G_{i \to j}$ displays singular power law tails (likely with logarithmic 
corrections) described by Eq.~(\ref{eq:tail}) with an exponent $1/2 < \mu < 1$, yielding a finite contribution
to the local DOS.   
The probability distribution of the real part (not shown) converges to 
a stationary distribution, whose tails are described by a power law (and logarithmic corrections) with
a $W$-independent exponent $1+\mu=2$ as in the fully localized phase.

\begin{figure}
\includegraphics[scale=0.305,angle=-90]{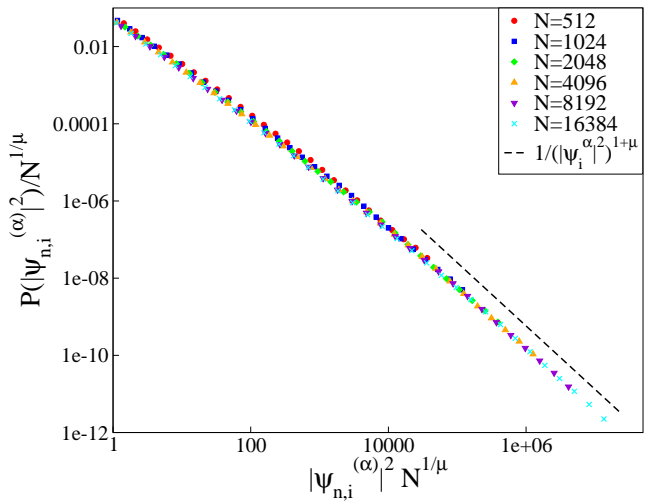}
\vspace{-0.2cm}
\caption{\label{fig:psi2}
Probability distribution of the wave functions amplitudes,
$P(|\psi_{n,i}^{(\alpha)}|^2)$, for $W=16$ and
different system sizes $N = 2^n$, with $n=9, \ldots, 14$, 
obtained from exact diagonalization of $2^{18-n}$ samples.
 The tails of the probability distributions are
described by a power law with an exponent $1+\mu=1.65$ (black
dashed lines) very close to the
the value of the exponent of the tails of $P(\textrm{Im}  G_{i \to j})$.
The tails of $P(|\psi_{n,i}^{(\alpha)}|^2)$ collapse on a single 
scaling function when plotted as a function of the scaling
variable $|\psi_{n,i}^{(\alpha)}|^2 N^{1/\mu}$.}
\vspace{-0.4cm}
\end{figure}

\subsection{Wave-functions statistics in the delocalized phase}
In this section we report some results on the behavior of the
wave-functions coefficients, $|\psi_{n,i}^{(\alpha)}|^2$, obtained
from exact diagonalization of finite systems for $W<W_L$.
Since $\lim_{\eta \to 0} \textrm{Im} G_{ii} = \pi \sum_n \delta
(E - E_n^{(\alpha)}) | \psi_{n,i}^{(\alpha)}|^2 $ it is natural to
expect that the rare and very large values of $\textrm{Im}  G_{ii}$
are related to values of $| \psi_{n,i}^{(\alpha)}|^2$ much larger than the 
typical ones (this is true only for $W<W_L$ where the spectrum is absolutely continuous, not for
$W>W_L$ where instead large values of $\textrm{Im}  G_{ii}$ are produced by localized rare
resonances). More precisely, we expect that power law tails for $\textrm{Im} G_{ii}$ should be related to
power law tails for  $P(|\psi_{n,i}^{(\alpha)}|^2)$.
For strong enough disorder, 
we indeed observe that the tails of $P(|\psi_{n,i}^{(\alpha)}|^2)$
are described by a power law. The exponent value changes (increases)
with the system size and approaches the value $\mu$ describing the tails of $P(\textrm{Im} G_{ii})$, as shown 
in Fig.~\ref{fig:psi2}.
We interpret this finite size effect in terms of fluctuations of the effective value of $W_L$
and the existence, for finite system sizes, of localized eigenstate even for $W<17.5$. 
These lead to spurious larger values of $|\psi_{n,i}^{(\alpha)}|^2$ that 
make the exponent of the tails effectively larger than its large-N limit. 

As previously shown in~\cite{monthus-garel} we have found that 
$|\psi_{n,i}^{(\alpha)}|^2$ display a multi-fractal behavior (more
details will be presented in a forthcoming paper~\cite{futurepaper}).
Multi-fractality implies that the probability distributions of the
wave-functions coefficients for different system sizes cannot be
fully rescaled in terms of the variable $|\psi_{n,i}^{(\alpha)}|^2
N^\kappa$ using a {\em unique} exponent $\kappa$. On the contrary, 
the value of $\kappa$ should depend on the range of $|\psi_{n,i}^{(\alpha)}|^2$
one focuses on. In particular, the value $\kappa_t$ 
describing the rescaling of the typical value of $|\psi_{n,i}^{(\alpha)}|^2$ 
is quite different from the ones, $\kappa_{es}$ and $\kappa_{el}$, describing the rescaling
of extreme (very small and very big) values of the wave-function
amplitudes. 
We have found both analytically and numerically (more
details will be presented in a forthcoming paper~\cite{futurepaper})
that in the intermediate phase the extreme tail of the distribution $P(|\psi_{n,i}|^2)$ 
are given by
\[
\frac{const}{N|\psi_{n,i}|^{2(1+\mu)}}
\] 
which means that $\kappa_{el}=1/\mu$ as shown in Fig.~\ref{fig:psi2} for $W=16$.
There are logarithmic corrections to the previous formula important to take into account 
to find that  $\langle \textrm{IPR} \rangle=N \int_0^1 \textrm{d} |\psi_{n,i}^{(\alpha)}|^2 \, 
P(|\psi_{n,i}^{(\alpha)}|^2) |\psi_{n,i}^{(\alpha)}|^4$ vanishes logarithmically in the intermediate phase. 
In agreement with the discussion above, we find that the whole curve cannot be 
rescaled using $|\psi_{n,i}^{(\alpha)}|^2 N^\kappa_{el}$.

\subsection{Computation of the ``free energy'' $\phi(s)$. }
Before explaining the numerical computation of $\phi(s)$ we 
shall briefly discuss its physical meaning, see~\cite{derrida,monthus-garel,futurepaper} for more details (in particular concerning the
difference between quenched and annealed averages). 
Denoting $e^{-Rfs}$ 
the contribution of a given path of length $R$, one can rewrite the 
sum~(\ref{eq:phi}) as an integral over all paths giving a contribution 
characterized by a value of $f$ between $ \tilde f$ and $\tilde f+d\tilde f$ times the number of such paths. 
By denoting the latter $\exp(R\Sigma(f))$, one ends up with the expression
\[
\sum_{{\cal P}}
\prod_{(i^\prime,i^{\prime \prime})\in {\cal P}} |G_{i^\prime \to 
i^{\prime \prime}}|^s=\int d\tilde f \exp(R[-s\tilde f+\Sigma(\tilde f)]) 
\]
The value of $\tilde f$ that dominates the integral 
for $R\to \infty$ depends on $s$. For small enough $s$,
one finds that the saddle point value of $\tilde f(s)$ is such that
$\Sigma(\tilde f(s))>0$. In this regime an exponential number of paths, $k'^R$ 
(with $k'=e^{\Sigma(\tilde f(s))}$), contributes to the sum. 
By increasing $s$, $\Sigma(\tilde f(s))$ decreases until the value $s^\star$ is reached. At this point 
the generalized entropy $\Sigma(\tilde f(s))$ vanishes; in conclusion $\phi(s)$ is related to the Legendre Transform of $\Sigma(f)$, more
precisely $\phi'(s)=-\Sigma(f(s))/s^2$, and allows one 
to find out whether a finite or an exponential number of paths contributes
to the sum in~(\ref{eq:DP}). 

In order to compute 
$\phi(s)$, on each edge of the lattice
(and for each finite value of the imaginary
regulator $\eta$) we introduce
the variable
\begin{equation}
y_{i \to j} = {\sum_{{\cal P}}}
\prod_{(i^\prime,i^{\prime \prime})\in {\cal P}} |G_{i^\prime \to
i^{\prime \prime}}|^s,
\end{equation}
where ${\cal P}$ are all directed paths of length $R$ originating
from site $i$, and $(i^\prime,i^{\prime \prime})$ are all the directed
edges (including $i \to j $) belonging to the path.
It is straightforward to derive the following exact recursion relation
for $y_{i \to j}$:
\begin{equation} \label{eq:y}
y_{i \to j} = |G_{i \to j}|^s \sum_{j^\prime \in
\partial i / j} y_{j^\prime \to i},
\end{equation}
where $|G_{i \to j}|^s$ can be computed using Eq.~(\ref{eq:recursion}).
Eqs.~(\ref{eq:recursion}) and~(\ref{eq:y}) can be interpreted
as an exact self-consistent equation for the joint probability distributions
$Q(G_{i \to j}, y_{i \to j})$, 
which can be computed by iteration 
using a population dynamics algorithm with arbitrary numerical
precision.
At each new $R$-th step of the iteration, we can compute $\phi (s,R,\eta)$
as the average value of the logarithm of $y_{i \to j}$ over the
distribution $Q(G_{i \to j}, y_{i \to j})$, divided by $Rs$, as in
Eq.~(\ref{eq:y}). In order to take the $R \to \infty$ limit, after about
$10^4$ iterations we extrapolate
the asymptotic value of $\phi(s,R,\eta)$ with a power law fit of the form:
$\phi (s,R,\eta) \simeq \phi(s,\eta) + c(s,\eta) /R^\zeta$ ($\zeta=1$
for all values of $W$, $s$, and $\eta$) where $c(s,\eta)$ is a constant 
which depends on $s$ and $\eta$.
Finally, in order to take the limit $\eta \to 0$, we computed
$\phi(s,\eta)$ for several values of $\eta$ from $10^{-2}$ to $10^{-6}$ 
and used a fit of the form
$\phi(s,\eta) = \phi(s) + A \eta^\delta$ 
(we find that $\delta \sim 1/4$ for all values of $W$ and $s$).
Note that the limits $\eta \to 0$ and $R \to \infty$ only commute when
$\phi (s)$ is finite: whenever $\phi(s)=0$ we find that the
constant $c (s,\eta)$ diverges as $\eta \to 0$.

\begin{figure}
\includegraphics[scale=0.31,angle=-90]{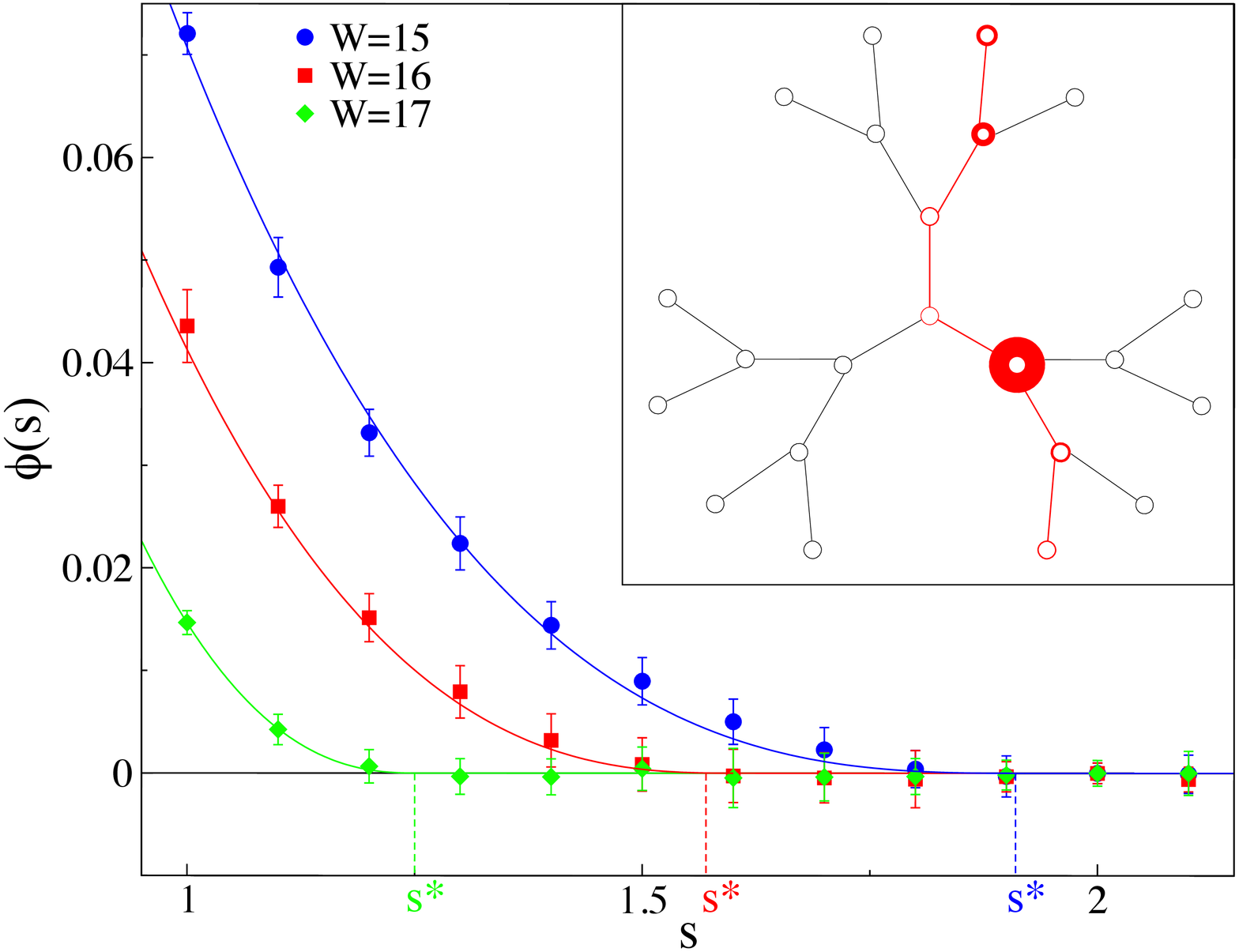}
\vspace{-0.2cm}
\caption{\label{fig:phi}
Main panel: Free energy, $\phi (s)$, as a function of $s$ in the
intermediate delocalized non-ergodic phase, for $W=15$ (blue circles), 
$W=16$ (red squares), and $W=17$ (green diamonds), showing the
value of $s^\star$ where $\partial \phi(s) / \partial s |_{s=s^\star} = 0$
and RSB occurs. Inset: Schematic sketch of the form of the wave-functions
in the delocalized non-ergodic phase.
Only few paths (the one colored in red) contribute significatively
to the value of $\textrm{Im}  G_{i \to j}$ on the central site.}
\vspace{-0.4cm}
\end{figure}

At weak enough disorder, in the delocalized ergodic phase, $W<W_T$, $\phi(s)$
is a smooth decreasing function of $s$. It is positive for $s<2$, decreases
as $s$ is increased and vanishes, with a positive derivative, at $s=2$. Hence, in this case the contribution
to $\textrm{Im}  G_{i \to j}$, Eq.~(\ref{eq:DP}), comes from an exponential
number of paths, since $\Sigma(f(2))>0$.
For $W=W_T$ the glass transition takes place: the derivative of $\phi(s)$ vanishes, {\it i.e.}~$s^{\star}=2$ (corresponding to
$\mu = 1$, {\it cfr.}~Fig.~\ref{fig:exp}) and $\Sigma(f(2))=0$.
In the intermediate delocalized non-ergodic phase, $W_T < W < W_L$,
$s^\star$ is found in the interval $(1,2)$: $\phi(s)$ is positive 
for $s<s^\star$, it vanishes at $s^\star$, and remains zero for higher values of $s$. Correspondingly,
the sum in~(\ref{eq:DP}) is dominated by a finite number of paths only.  
In Fig.~\ref{fig:phi} we show plots of $\phi(s)$ 
in the intermediate delocalized non-ergodic phase, 
showing the value of $s^\star$, which coincides (within the
numerical accuracy) with $2 \mu$ ({\it cfr.}~Fig.~\ref{fig:exp}).
Note that in agreement with~\cite{aizenmann}, we did find $s^\star = 1$ at $W_L$. 

\subsection{Energy transport.}
It is very important to figure out what happens to energy transport and 
diffusion in the new intermediate delocalized-ergodic phase.
In order to do that one can study the propagation of dissipation, {\it} i.e. of $\textrm{Im}  
G_{i \to j}$, on an open Bethe lattice where a thermal and particle reservoir is attached to the boundary sites. 
An approximate way to perform this analysis corresponds to study the recursion equations, 
Eq.~(\ref{eq:recursion}), where the imaginary regulator $\eta$
is set to a finite value of $O(1)$ only at the first step of the iteration,
and $\eta = 0$ for all the successive iterations. Each iteration
step corresponds to moving a generation further towards the bulk of the graph~\cite{abou}.

In the delocalized ergodic phase, $W<W_T$, after a certain number
of iterations $P (\textrm{Im} 
G_{i \to j})$ reaches a stationary distribution which coincides with 
the $\eta$-independent limit shown in Figs.~\ref{fig:tails} 
and~\ref{fig:limits}, as expected, and
$\langle \textrm{Im} 
G_{i \to j} \rangle$ has a well-defined value in the bulk.

In the fully localized phase, $W>W_L$, $P (\textrm{Im} 
G_{i \to j})$ does not have a stationary distribution. As we move
deeper towards the center of the graph, the whole distribution globally shifts
exponentially towards smaller values of $\textrm{Im} 
G_{i \to j}$. Both $\langle \textrm{Im} 
G_{i \to j} \rangle$ and $\exp(\langle \log \textrm{Im} 
G_{i \to j} \rangle)$ go exponentially to $0$ as the number of iteration
is increased. As expected, the system, even if coupled at the boundary to a reservoir, 
does not show any dissipation and decoherence in the bulk. 

In the delocalized non-ergodic phase, $W_T < W < W_L$, the situation
is intermediate between the two cases described above. As exlained in the main text,
the fact that the sum in~(\ref{eq:DP})
is dominated by few paths only implies that particle and energy propagation
proceeds over few specific disorder dependent paths. Therefore, 
there is dissipation and decoherence in the bulk but it
is extremely heterogeneous. 
Indeed, we find that 
the energy of the reservoir at the boundary penetrates towards the center of the system, 
since on almost all the sites of the bulk $\textrm{Im} 
G_{i \to j}$ is finite, albeit possibly very small, as demonstrated by the fact that 
$\exp(\langle \log \textrm{Im} G_{i \to j} \rangle)$ approaches a (very small) finite value.
However, there are very rare resonances
where $\textrm{Im} G_{i \to j}$ is huge (possibly corresponding to the singular power law tails
of $P (\textrm{Im} G_{i \to j})$ found when $\eta$ is different from zero on all sites). 
The behavior of such extreme events should be better characterized and understood. 
Qualitatively, we observe that
the deeper we move towards the bulk of the lattice the rarer
and bigger the resonances become.


\begin{thebibliography}{25}
\bibitem{footnotetree}
Localization was studied on both Cayley trees and Bethe lattices. The formers might display some spurious behavior 
due to the peculiarity of having a finite fraction of sites on the boundary, as shown in~\cite{berkovits} for the level statistics. Whether the delocalized non-ergodic phase found in this work for Bethe lattices also shows up for Cayley trees is a question worth addressing.
\bibitem{footnoteWT}
For $W=W_T$ we expect $P(r)$ to converge to a universal distribution different from GOE and Poisson.
\bibitem{footnoteAi}
Note that in~\cite{aizenmann} instead of the average of the logarithm it was considered the logarithm of the average. 
These are called quenched and annealed averages in DPRM terminology. They coincide for $s\le s^\star$ and, hence,
can be equivalently used to find out the value of $s^\star$.
\bibitem{footnote2}
The analysis in~\cite{fyodorovsparse} 
assumed the existence of the second derivative of the order parameter function, which coincides with 
$\langle (\textrm{Im} G_{ii})^2 \rangle$. Since $\mu$ becomes smaller than two before $W_L$ (and also before $W_T)$
this hypothesis does not hold except for very small disorder.
\end{thebibliography}
\end{document}